\documentstyle[preprint,floats,tighten,epsf,prd,aps]{revtex}

\def\be{\begin{equation}}
\def\ee{\end{equation}}
\def\half{{1\over 2}}
\def\bea{\begin{eqnarray}}
\def\eea{\end{eqnarray}}
\def\bml{\begin{mathletters}}
\def\blea{\begin{mathletters}\begin{eqnarray}}
\def\elea{\end{eqnarray}\end{mathletters}}

\def\ba{{\bf a}}
\def\bb{{\bf b}}
\def\bc{{\bf c}}
\def\bx{{\bf x}}
\def\bbeta{{\bbox{\beta}}}

\def\betapar{{\beta_\parallel}}

\def\xdot{\dot{\bf x}}
\def\atilde{{\tilde{\bf a}}}
\def\btilde{{\tilde{\bf b}}}
\def\ttilde{{\tilde t}}

\def\jtilde{{\tilde \jmath}}
\def\Atilde{{\tilde A}}
\def\Btilde{{\tilde B}}
\def\mutilde{{\tilde\mu}}

\begin{document}
\long\def\ignore#1{}
\draft
\title{A vorton gun}

\author{Ken D.\ Olum\footnote{Email address: {\tt kdo@alum.mit.edu}},
J.\ J.\ Blanco-Pillado\footnote{Email address: {\tt
jose@cosmos2.phy.tufts.edu}},
and Xavier Siemens\footnote{Email address: {\tt
siemens@cosmos2.phy.tufts.edu}}}

\address{Institute of Cosmology \\
Department of Physics and Astronomy \\
Tufts University \\
Medford, MA 02155}

\date{September 2000}

\maketitle

\begin{abstract}%
In about half of all near-cusp events in superconducting cosmic
strings with chiral currents (and probably with general currents as
well), the string intersects itself near the cusp.  Intercommutation
causes the conversion of the string near the cusp into a vorton (in
the chiral case) with very high Lorentz boost.  We demonstrate how to
analyze the cusp shape in a Lorentz frame that makes the motion
simple, by the use of a 5-dimensional procedure, and analyze the
resulting production of vortons.
\end{abstract}

\narrowtext

\section{introduction}
Cosmic strings are topological defects that may have been created by
symmetry breaking phase transitions in the early universe.  (For
reviews see\cite{Alexbook,Kibble95}).  Witten\cite{Witten85}
discovered that in many theories, cosmic strings could acquire a
condensate of zero-mass particles, and thus could behave as
superconducting wires carrying charge or current.  The existence of
charges and currents on the string can substantially modify its
dynamics.  If the charge and current are large enough, the string may
have a stable static configuration known as a vorton
\cite{Davis88-2}.

For a non-superconducting string, as long as the string thickness
(typically a tiny microphysical quantity) is small as compared to its
radius of curvature (typically of cosmological size), the motion of
the string is given by the Nambu-Goto equations of motion, which are
easily solved.  With an appropriate choice of parameter (gauge), the
string position can be given by a function $\bx (\sigma, t)$, where
$\sigma$ parameterizes the energy on the string, and $t$ is the
ordinary time coordinate.  The general motion of the string can then
be written\footnote{We work in units where $c = 1$.}
\be\label{eqn:motion}
\bx (\sigma, t) =\half\left[\ba (\sigma -t) +\bb (\sigma +t)\right]
\ee
where the functions $\ba $ and $\bb$ are arbitrary except that they
must obey the constraints
\blea
|\ba' (\sigma) |^2 & = & 1\\
|\bb'  (\sigma) |^2 & = & 1\,.
\elea

In general, a Nambu-Goto string loop will have two or more points at
which $\bb' (\sigma, t) = -\ba' (\sigma, t)$.  Such a point is called
a cusp.  It has $\bx' = (\ba' +\bb')/2 = 0$ and $\xdot = (\bb' -\ba')/2
=\bb'$.  Thus $ |\xdot | = 1$; a point moves at the speed of light.

For a superconducting string, the equations of motion are much more
complicated.  However, if we can neglect long-range electromagnetic
interactions (either because the condensate is not coupled to the
electromagnetic field, or because such corrections are small), the
situation is much simpler.  If furthermore the current is ``chiral''
(charge and current equal in magnitude), then the equations can be
solved exactly\cite{Carter99-1,Blanco-Pillado00-1,Davis00-1}.  This
situation can occur both in strings which can carry currents in only one
direction\cite{Davis97,Carter99-2}, but also in more usual
superconducting strings.  As the string approaches a cusp, all
charge-carriers will be accelerated, and left-moving and right-moving
charge carriers can scatter with each other and be ejected from the
string.  This will lead to a predominance of whichever type of
charge-carrier was in larger numbers originally, and thus an approach
to a chiral state.

In the chiral case, the motion is still given by Eq.\
(\ref{eqn:motion}), but now the constraints are
\bml\label{eqn:newconstraints}\bea
|\ba'  (\sigma) |^2 & = & 1\\
|\bb'  (\sigma) |^2 & = & 1-j^2
\elea
(or the same with $\ba$ and $\bb$ exchanged, depending on the
direction of current flow),
where $j$ is a measure of the magnitude of the current, to be
discussed later.  From Eqs.\ (\ref{eqn:motion}) and
(\ref{eqn:newconstraints}) we can see that such a string
cannot have a cusp, per se.  Since $\ba' $ and $\bb' $ have different
magnitudes, they cannot cancel exactly, and since $|\bb'| < 1$, we
cannot have $|\xdot | = 1$.

However, if $j\ll 0$, then we can have an event that looks very
similar to a Nambu-Goto cusp at large distances, but has a different
structure near the place that the cusp would have occurred.  Thus
we will consider a point with $\ba' $ and $\bb' $ pointing in opposite
directions, even though their magnitudes are different, so $\bb' = -
\sqrt{1-j^2}\ba'$.  This means that $\bx'$ has some small but nonzero
magnitude.  We will show that this can lead to a self-intersection
near the cusp and the emission of a vorton.

\section{self intersection}
Because the two-dimensional world sheet current is conserved, we can
describe it as the curl of an auxiliary scalar field $\phi$.  The
world sheet current is then
\be\label{eqn:current}
J^a = q {1 \over{\sqrt{-\gamma}}} \,\epsilon^{ab} \, \phi_{,b}
\ee
where $q$ is the electric charge multiplied by the integral of the
field density across the string core, and renormalized if necessary,
and $\gamma^{ab} $ is the induced metric in the world sheet.
The physical current is
\be
J^\mu (x) =\int d\sigma dt\sqrt{-\gamma} J^ax^\mu_{,
a}\delta^{(4)}(x-x (\sigma,t)) = q\int d\sigma\epsilon^{ab}
\phi_{, a} x^\mu_{, b}\delta^{(3)} (\bx-\bx (\sigma, t))
\ee

In the chiral case, we can assume without loss of generality that the
charge carriers move in the negative $\sigma$ direction, so that
$\phi$ is a function only of $\sigma +t$, and thus $\phi '
=\dot\phi$.  We will define $j = 2\phi'/\sqrt{\mu}$, which gives the
constraints of Eqs.\ (\ref{eqn:newconstraints}).  The current $j$ is
dimensionless and gives a measure of the ratio of the energy in the
current to the energy in the string.  With $j = 1$ we would have $\bb'
= 0$ which gives a static vorton solution.

We will consider a cusp in a string loop with $j\ll1$, and we
will define $\Delta = |\bb' | =\sqrt{1-j^2}$ and $\epsilon =
1-\Delta$, and let subscript $0$ denotes quantities at the time and
place of the cusp.  We have $|\ba'_0 | = 1$, $\bb'_0 = -\Delta\ba'_0
$, and so
\blea
\bx'_0 & =& (\ba'_0 +\bb'_0)/2 = (\epsilon/2)\ba'_0\\
\xdot_0 & = & (\bb'_0-\ba'_0)/2 = - {1+\Delta\over 2}\ba'_0
\elea

In\cite{jjkdo98.0} we showed that, for an ordinary string cusp, one can
choose a Lorentz frame in which $\xdot_0$ and $\bx'''_0$ are parallel.
In the appendix, we extend this result to chiral superconducting
strings, and show that we can work in a frame with $\bx'_0$ and
$\bx'''_0$ parallel.  In that frame, we can expand the position of the
string at the time of the cusp.  We take the cusp to occur at $\sigma
= t = 0$ and its position to be $\bx = 0$.  Thus
\be
\bx (\sigma) =\bx'_0\sigma +\bx''_0{\sigma^2\over 2} +\bx'''_0{\sigma^3\over 6}
+\cdots
\ee
If $\bx'_0$ and $\bx'''_0$ point in opposite directions, which one
would expect roughly half the time, then the string will intersect
itself at $\sigma =\pm\sigma_1$, where
\be
\sigma_1 =\sqrt{|\bx'_0 |\over6 |\bx'''_0 |}
\ee
This self-intersection will lead to an intercommutation and the
emission of a loop containing the original string from
$-\sigma_1$ to $\sigma_1$.

\section{the emitted loop}
After the intercommutation, the main string will have kinks where the
cusp would have been, and there will be a small loop (also with kinks) that
has been produced by intercommutation.  The effect on the main string
is very similar to emission of string due to overlap of the core, as
discussed in \cite{jjkdo98.1}.  Here we will study the small loop that
is emitted.

First of all, what amount of the string is emitted?  The magnitude of
$\bx'_0$ is $\epsilon/2$, while $\bx'''_0$ could be expected to be of
order $L^{-2}$, where $L$ is the size of the original loop, or the
typical feature size in the case of a long string.  Thus we
expect $\sigma_1 = O (\sqrt{\epsilon}L) = O (jL)$.  If we start out,
for example, with 1\% of the vorton current, then roughly 1\% of the
original loop will be chopped off.

We would like
to compute the velocity and rest-frame properties of the emitted loop.
The energy-momentum tensor is given by\cite{Blanco-Pillado00-1}
\be
T^{\mu\nu} =\int d\sigma d\tau\sqrt{-\gamma}\left(\gamma^{ab}
+\theta^{ab}\right) x^\mu_{,a} x^\nu_{,b}\delta^4 [x-x (\sigma,\tau)]
=\mu\int d\sigma \eta^{ab} x^\mu_{, b} x^\nu_{,a}\delta^3 [\bx-\bx (\sigma,t)]
\ee
and thus the total energy-momentum in a region is
\be
P^\mu =\int d^3xT^{\mu 0} =\mu\int d\sigma \eta^{ab} x^\mu_{, b}x^0_{,a}
=\mu\int d\sigma \dot x^\mu
\ee
Since $x^0 = t$, we have $\dot x^0 = 1$, and so the emitted energy is
just $E = P^0 = 2\mu\sigma_1$, which is just the energy in a segment
of string of length $2\sigma_1$.  We can expand $\xdot$
in a Taylor series,
\be
\xdot =\xdot_0+\sigma\xdot'_0+\half\sigma^2\xdot''_0
\ee
and integrate to find the momentum,
\be
{\bf P} = 2\mu\sigma_1\xdot_0+{\sigma_1^3\over 3}\mu\xdot''_0
= E \left(\xdot_0+{\sigma_1^2\over 6}\xdot''_0\right)=
E\left(-\ba'_0+ {\epsilon\over 2}\ba'_0+{\sigma_1^2\over 6}\xdot''_0\right)\,.
\ee
The magnitude of $\ba'_0$ is 1, whereas the magnitude of the second
term is $O (\epsilon)$, and the third is likewise $O (\epsilon)$, since $\sigma_1^2 = O
(\epsilon L^2)$ and $\xdot''_0 = O (L^{-2})$.  Furthermore, the second
and third terms cannot cancel, because
 $\xdot''_0\cdot\ba'_0 = (\bb'''_0 -\ba'''_0)\cdot\ba'_0/2 = O (|\bb''_0
|^2) + O (|\ba''_0 |^2) > 0$, so the two vectors cannot point in
opposite directions.  Thus the magnitude of $\bf P$ is
\be
P = E [1-O(\epsilon)]\,.
\ee

The rest-frame energy of the loop is $\sqrt{E^2-P^2}$, so the length
of the loop in its rest frame is
\be
\tilde\sigma_1 =\mu^{-1}\sqrt{E^2-P^2} =\mu^{-1} E\, O (\sqrt{\epsilon}) = 2\sigma_1 O (j) = O (\epsilon L)
\ee
Since $\phi$ is a scalar, it is unaffected by the boost. The current
$j$ is proportional to $\phi' = d\phi/d\sigma$ and thus is enhanced by
the ratio of $\sigma$ in the moving frame to $\sigma$ in the rest
frame, which is $O (1/j)$.  The new value is $\jtilde = O (1)$,
so we have essentially a vorton.

What is the boost vector $\bbeta$ that brings the loop to rest?  It
must point in the direction of $-\xdot_0$ and should give a Lorentz
factor $\gamma =\sigma_1/\tilde\sigma_1 = O (\epsilon^{-1/2}) = O
(j^{-1})$.  Thus the boost should have magnitude $\beta
=\sqrt{1-\gamma^{-2}} =1-O (\epsilon)$, so $\bbeta\sim-\bb'_0$.  Then
from Eq.\ (\ref{eqn:fBdef}), $f_B\sim1/(\gamma (1-\beta^2))
=\gamma\sim1/j_0 $.  From Eq.\ (\ref{eqn:jtransform}) the current at
the cusp transforms as
\be
\jtilde_0 =  f_Bj_0\sim 1\,,
\ee
in accord with the above, and from Eq.\ (\ref{eqn:j2transform}) the derivative of the current is
\be
\jtilde'_0 = f_B^2j'_0+f_Bf_{B, B} j_0 \sim {j'_0\over j_0^2}
+{\bbeta\cdot\bb''_0\over j_0^3}\,.
\ee 
Now $\bbeta\cdot\bb''_0 = -j'_0 j_0 $, so the second term is of the same
order as the first, and $\jtilde'_0\sim j'_0/j_0^2\sim 1/(j_0 L)$.
This means that the change in  current over the loop is
\be
\jtilde (\tilde\sigma_1) -\jtilde_0\sim (1/(j_0L))\cdot j_0^2L = j_0
\ll\jtilde_0\sim 1
\ee
and thus the current is essentially constant over the loop.

\section{discussion}

\begin{figure}
\begin{center}
\leavevmode\epsfbox{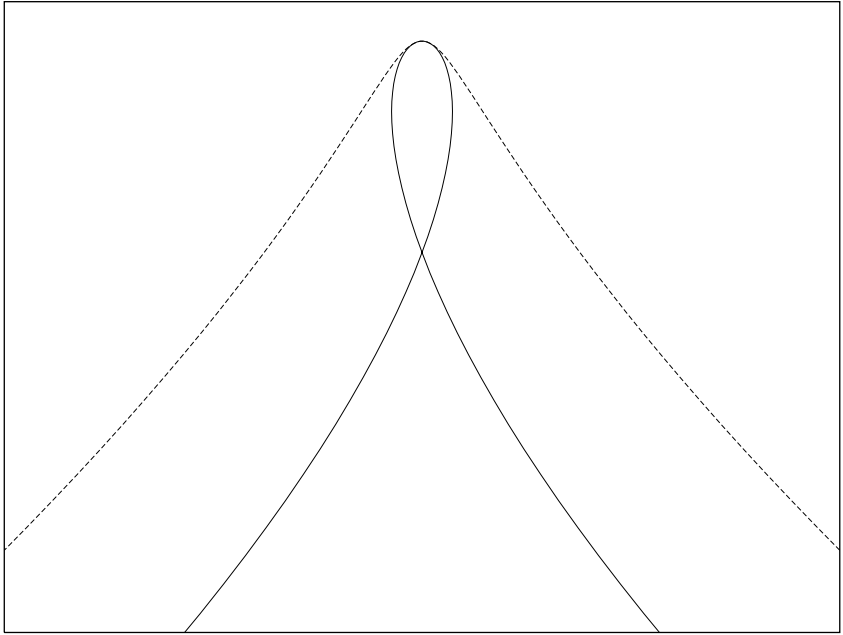}
\end{center}
\caption{Cusp-like events in a chiral superconducting string.  Dashed
line: $\bx'_0$ and $\bx'''_0$ parallel.  Solid line: $\bx'_0$ and $\bx'''_0$ antiparallel.}
\label{fig:cusps}
\end{figure}

We have shown that cusps (or, perhaps more properly, ``cusp-like
events'') in chiral strings come in two forms.  If $\bx'''_0$ points in
the same direction as $\bx'_0$, in the frame in which these are
parallel, then the cusp is merely smoothed out, as shown by the dashed
line in Fig.\ \ref{fig:cusps}.  However, if $\bx'''_0$ points in the
opposite direction to $\bx'_0$, then there is a self intersection, as
shown by the solid line in Fig.\ \ref{fig:cusps}.  We thus expect that
half of all cusps in strings with chiral currents will have self
intersections.  When the string does intersect itself,
intercommutation emits a loop of string with large Lorentz factor
which is approximately in a vorton state already, and could be
expected to become a stable vorton.

What about currents which are not chiral?  We do not have an exact
solution in this case, but we expect similar behavior.  The effect of
the current is still to prevent the string from doubling back on
itself, so there will be a finite value of $\bx'_0$ where the point of
the cusp would have been.  This vector controls the shape very close
to this point.  Further away, however, the effect of $\bx'''_0$ becomes
more important, and if these vectors point in opposite directions, we
might expect a self intersection.

If a system of strings can produce vortons, then it is very likely
that the vortons will contribute to the matter density of the universe
in amounts excess of observation.  Thus most such theories are ruled
out\cite{Brandenberger:1996zp,Carter99-2}.  For a string to become a
vorton, the current must somehow rise to the necessary level to oppose
the tension in the string.  We have shown here a new mechanism by
which strings with small currents can nevertheless produce stable
vortons.  This mechanism may lead to vorton production in more
theories than would otherwise have this problem, and thus restrict
further the range of allowable superconducting string theories.

It should be noted that, in the case of the string with chiral current
and no coupling to the electromagnetic field, the motion will be
strictly periodic\cite{Blanco-Pillado00-1}.  After a single cusp
has formed at a particular point in the oscillation of the string
loop, and a vorton has been emitted, there will be a kink, rather
than a cusp, and no further emission of vortons.  However, long
strings, currents that are not exactly chiral, and coupling to the
electromagnetic field produce motion which is not exactly periodic,
and thus perhaps lead to ongoing vorton emission.  Smoothing of the
kink may also produce future cusps, as discussed in\cite{jjkdo98.1}.

\appendix
\section{Lorentz transformations}
In this appendix we extend our results\cite{jjkdo98.0} for Lorentz
boosting of ordinary string cusps to superconducting strings with
chiral currents.

The motion of a chiral superconducting string is given by Eq.\
(\ref{eqn:motion}) with the constraints of Eqs.\
(\ref{eqn:newconstraints}).  Following the technique of
\cite{Nielsen:1987fy,Nielsen:1980zf}, we can treat the current in the
same way as the physical position by considering a 4+1-dimensional
spacetime where the fourth spatial component of $\bb$ is $b_4 =
2\phi/\sqrt{\mu}$, so that the fourth spatial component of $\bb'$ is
$j$.  Since the current travels only in one direction, the new
components of $\ba$ and its derivatives are all 0.

In this appendix, we will let $\ba_3$ and $\bb_3$ denote the 3-space
vectors previously called $\ba$ and $\bb$, let $\ba$ and $\bb$
denote the new 4-space vectors, and introduce 5-dimensional
spacetime vectors $A$ and $B$.

In general, we have
\blea
\ba' & = & (\ba'_3,0)\\
\bb' & = & (\bb'_3, j)\\
\ba'' & = & (\ba''_3,0)\\
\bb'' & = & (\bb''_3, j')
\elea
and so on.  At the cusp
\be
\bb' = (- \Delta\ba'_3, j)\,.
\ee

Each 4-vector has the same relationship with its own derivatives as
usual,
\blea
\ba'' \cdot\ba' & = & 0\\
\ba'''\cdot\ba' & = & - |\ba'' |
\elea
and likewise for $\bb$.

At the cusp, we can take $\bb'_3$,  which is
parallel to $-\ba'$, as the ``direction of the cusp''.  The vector $\ba''$ is
perpendicular to this vector, but $\bb''_3$ will not be perpendicular.
Instead we have, at the cusp,
\be
0 = \bb''\cdot\bb' = - \Delta\bb''\cdot\ba'+jj'
\ee
so
\be
\bb''_3\cdot\ba'_3 =\bb''\cdot\ba'  = {jj'\over \Delta}\,.
\ee
We can define a new vector $\bc =\bb''_3-( jj'/\Delta)\ba'$, so that
\be
\bc\cdot\ba' = 0\,.
\ee 
Now we have two vectors, $\ba''$ and $\bc$, in
general independent, which are both perpendicular to $\bx'_3$.  Both
have vanishing component in the current direction.  Thus
if we can find a frame where $\bx''' $ is perpendicular to both $\ba''$
and $\bc$, then $\bx'''_3$ must be parallel to $\bx'_3$.

The remaining inner products are given by
\blea
- |\ba'' |^2 & = &\ba'''\cdot\ba' = -{\ba'''\cdot\bb'\over \Delta}\\
- |\bb'' |^2 & = &\bb'''\cdot\bb' = - \Delta\bb'''\cdot\ba'
+j'' j
\elea
so
\blea
\ba'\cdot\bb' & = & - \Delta\\
\ba''\cdot\bb' & = & 0\\
\ba'''\cdot\bb' & = & \Delta
|\ba'' |^2\\
\bb'''\cdot\ba' & = & {|\bb'' |^2\over \Delta} +{jj''\over
\Delta}\\
\elea

As in\cite{jjkdo98.0}, we have two null spacetime vectors, here 5-dimensional,
\blea
A^\mu & = & (1, -\ba')\\
B ^\mu & = &  (1,\bb')\,.
\elea
We define
\blea
B_2^\nu & = & B ^\mu\partial_\mu B ^\nu = (0,\dot\bb' +\bb'') = (0,2\bb'')\\
A_2^\nu & = & A ^\mu\partial_\mu A ^\nu = (0,-\dot\ba' +\ba'') = (0,
2\ba'')\\
A_3 ^\rho & = & A ^\mu\partial_\mu (A ^\nu\partial_\nu A ^\rho)=(0, -4\ba''')\\
B_3 ^\rho  & = &B ^\mu\partial_\mu (B ^\nu\partial_\nu B ^\rho)=(0, 4\bb''')
\elea
and we have inner products at the cusp,
\blea
g (A, B) & = &\epsilon\\
g (A_2, A) & = & g (B_2, B) = 0\\
g (A_2, B) & = & 0\\
g (B_2, A) & = &{2jj'\over \Delta}\\
g (A_3, A) & = & -4 |\ba'' |^2\\
g (B_3, B) & = & -4 |\bb'' |^2\\ 
g (B_3, A) & = & {4 |\bb'' |^2\over \Delta} +{4jj''\over
\Delta}\\
g (A_3, B) & = &4 \Delta |\ba'' |^2 \,.
\elea

As before, a coordinate transformation yields
\blea
A ^\mutilde & = & (A ^ \ttilde, {\bf A})\\ B ^\mutilde & = & (B ^ \ttilde, {\bf
B})\,.
\elea
and
\blea
\Atilde ^\mutilde &= & (1,-\atilde') = A/A ^ \ttilde\\
\Btilde ^\mutilde  &= & (1,\btilde') = B/B ^ \ttilde\,.\label{eqn:Btransform}
\elea
We will consider only boosts which do not involve the fifth direction.
The above then says that $\bb'_3$ is boosted and rescaled by the
transformation, while $j $ is merely rescaled, because there is
no boost in that direction.

We will let our new coordinate system move with velocity $-\bbeta$, so
that a particle at rest in the original system is moving with velocity
$\bbeta$ with respect to the new coordinates.  The vector $\bbeta$ is
a space 4-vector, but has zero current component.  The Lorentz
transformation then gives $A ^\ttilde =\gamma (1 -\ba'\cdot\bbeta)$
and $B ^\ttilde =\gamma (1 +\bb'\cdot\bbeta)$, where $\gamma = 1/\sqrt
{1-\beta ^ 2}$.  We will define
\blea
f_A & = & {1\over A ^\ttilde} = {1\over\gamma (1-\ba'\cdot\bbeta)}
\label{eqn:fAdef}\\
f_B & = & {1\over B ^\ttilde} = {1\over\gamma (1+\bb'\cdot\bbeta)}
\label{eqn:fBdef}
\elea
exactly as before.
From Eq.\ (\ref{eqn:fBdef}) and (\ref{eqn:Btransform}) we see that the
current transforms as
\be\label{eqn:jtransform}
\jtilde = f_Bj\,.
\ee

We find the transformation laws
\blea
\Atilde_2 &= &f_A ^ 2A_2 + f_Af_{A, A} A\\
\Btilde_2 &= & f_B ^ 2B_2 + f_Bf_{B, B} B\label{eqn:B2transform}\\
\Atilde_3  &= & f_A ^ 3 A_3 + 3f_A ^ 2f_{A, A} A_2 + (f_Af_{A, A} ^2 + f_A^2 f_{A, AA}) A\\
\Btilde_3  &= & f_B ^ 3 B_3 + 3f_B ^ 2f_{B, B} B_2 + (f_Bf_{B, B} ^ 2 + f_B^2 f_{B, BB}) B
\elea
with $f_{A, A} = A ^\mu\partial_\mu f_A$ and $f_{B, B} = B
^\mu\partial_\mu f_B$, $f_ {A, AA} = A ^\mu\partial_\mu (A
^\nu\partial_\nu f_A)$ and $f_ {B, BB} = B ^\mu\partial_\mu (B
^\nu\partial_\nu f_B)$.
We also have
\blea
f_{A, A}  & = & -2\gamma f_A ^ 2\bbeta\cdot\ba''\\
f_{B, B} & = & -2\gamma f_B ^ 2\bbeta\cdot\bb''\,.
\elea
Applying Eq.\ (\ref{eqn:B2transform}) to the derivative of the
current, we find in particular,
\be\label{eqn:j2transform}
\jtilde' = f_B^2j'+f_Bf_{B,B} j\,.
\ee

Every result from\cite{jjkdo98.0} that is not specific to the point
of the cusp goes through unchanged.  
Thus
\blea
|\atilde''| & = & f_A ^ 2 |\ba''|\\
|\btilde'' | & = & f_B ^ 2 |\bb'' |
\elea
as before.  Note, however, that
$\bb''$ includes $b''_4 = j' $. 

For the mixed product, however, things are more complicated.  We have,
at the cusp,
\be
\atilde''\cdot\btilde'' = f_A^2f_B^2\ba''\cdot\bb''
+f_B^2f_Af_{A, A} {jj'\over 2 \Delta}+{\epsilon\over 4} f_Af_{A, A} f_Bf_{B, B}\,.
\ee 

The same situation applies to the third derivatives.  For all A's or B's,
everything is the same, and we get
\blea
\atilde'''\cdot\atilde'' & = & f_A ^ 5\ba'''\cdot\ba'' - f_A ^ 4f_{A, A}
|\ba''| ^ 2
= f_A ^ 5 (\ba'''\cdot\ba''+ 2\gamma f_A
|\ba''| ^ 2\bbeta\cdot\ba'')\\
\btilde'''\cdot\btilde'' & = & f_B ^ 5\bb'''\cdot\bb'' + f_B ^ 4f_{B, B}
|\bb''| ^ 2
= f_B ^ 5 (\bb'''\cdot\bb'' -2\gamma f_B |\bb''|
^ 2\bbeta\cdot\bb'')
\elea
For cross terms, we have
\blea
-8\atilde'''\cdot\btilde'' & = & g (\tilde A_3,\tilde B_2)
= f_A^3f_B[f_Bg (A_3, B_2) +f_{B, B} g (A_3, B)]
 +3f_A^2f_{A, A} f_B[f_Bg (A_2, B_2)\nonumber\\
& & + f_{B, B} g (A_2, B)]
+(f_Af_{A, A} ^2 + f_A^2 f_{A, AA}) f_B[f_Bg (A, B_2) +f_{B, B} g (A,
B)]\\
& = &f_A^3f_B[-8f_B\ba'''\cdot\bb''+4f_{B, B} \Delta |\ba'' |^2]
 +12f_A^2f_{A, A} f_B^2\ba''\cdot\bb''\nonumber\\
& & +(f_Af_{A, A} ^2 + f_A^2 f_{A, AA}) f_B[2f_Bjj'\Delta^{-1} +f_{B,
B}\epsilon]\\
8\btilde'''\cdot\atilde'' & = & g (\tilde B_3,\tilde A_2)
= f_B^3f_A[f_Ag (B_3, A_2) +f_{A, A} g (B_3, A)]
 +3f_B^2f_{B, B} f_A[f_Ag (B_2, A_2)\nonumber\\
& & + f_{A, A} g (B_2, A)]
+(f_Bf_{B, B} ^2 + f_B^2 f_{B, BB}) f_A[f_Ag (B, A_2) +f_{A, A} g (B,
A)]\\
& = &f_B^3f_A[8f_A\bb'''\cdot\ba'' +4\Delta^{-1} f_{A, A} |\bb''  |^2
+4\Delta^{-1} jj'']\nonumber\\
& & +3f_B^2f_{B, B} f_A[4f_A\ba''\cdot\bb'' +2f_{A, A}\Delta^{-1} jj']
+(f_Bf_{B, B} ^2 + f_B^2 f_{B, BB}) f_A f_{A, A}\epsilon
\elea

Now we consider the case that $\epsilon\ll 1$, and so $\Delta = 1+O
(\epsilon)$ and $j^2 = O (\epsilon)$.  We will look for a Lorentz
transformation where $f_B = 1$, so the current is unchanged by the
boost. 
This means that $1-\Delta\betapar = 1-\betapar +\epsilon\betapar =
1/\gamma$, and so
\be
f_A ={1 -\Delta\betapar\over 1-\betapar} = 1+{\epsilon\betapar\over
1-\betapar}= 1+O (\epsilon) \,.
\ee

There are three parameters in $\bbeta$, but we have fixed one of them
by demanding $f_B = 1$.  This leaves two parameters of the Lorentz
transformation, which we can vary in an effort to solve the two
simultaneous equations $\tilde\bx'''\cdot\tilde\ba'' = 0$ and
$\tilde\bx'''\cdot\tilde\bc = 0$.  These are the same equations that
we solved in\cite{jjkdo98.0} for the ordinary string case, with
additional perturbations of order $\epsilon\ll1$. For generic values
of the cusp parameters, the solution will be stable and thus the
perturbed equations will still be solvable and the effect of the
perturbation will be to change the solution by terms of order
$\epsilon$.  Thus in the present case we will be able to find a frame
in which $\bx'''_3$ and $\bx'_3$ are parallel, and,  as
in\cite{jjkdo98.0}, the required boost will not be particularly large.


\end{document}